\documentclass[12pt,preprint]{aastex}

\def\Prob   {\ifmmode {{\rm Prob}} \else {Prob} \fi}
\def\arm    {\ifmmode {{\rm arm}} \else {arm} \fi}

\def\twopi  {\ifmmode {2\pi}   \else {$2\pi$}   \fi}

\def\SgrA {Sgr~A*}

\def\kms    {\ifmmode{{\rm km s}^{-1}}\else{km s$^{-1}$}\fi}

\def\uas  {\ifmmode {\mu{\rm as}}\else{$\mu$as}\fi}
\def\deg  {\ifmmode {^\circ}\else {$^\circ$}\fi}
\def\porm {\ifmmode {\pm}\else {$\pm$}\fi}

\def\chisqpdf {\ifmmode {\chi^2_{\rm pdf}}\else {$\chi^2_{\rm pdf}$}\fi}
\def\chisq    {\ifmmode {\chi^2}\else {$\chi^2$}\fi}

\def\HI   {H~{\small I}}

\def\HII  {H~{\small II}}

\def\etal {et al.~}
\def\eg   {e.g.,~}

\def\d    {\ifmmode {{\rlap{.}}^\circ}\else {${\rlap{.}}^\circ$}\fi}
\def\s    {\ifmmode {{\rlap{.}}^s}\else {${\rlap{.}}^s$}\fi}
\def\as   {\ifmmode {{\rlap{.}}^{''}}\else {${\rlap{.}}^{''}$}\fi}

\newbox\grsign \setbox\grsign=\hbox{$>$} \newdimen\grdimen \grdimen=\ht\grsign
\newbox\laxbox \newbox\gaxbox
\setbox\gaxbox=\hbox{\raise.5ex\hbox{$>$}\llap
     {\lower.5ex\hbox{$\sim$}}}\ht1=\grdimen\dp1=0pt
\setbox\laxbox=\hbox{\raise.5ex\hbox{$<$}\llap
     {\lower.5ex\hbox{$\sim$}}}\ht2=\grdimen\dp2=0pt

\def\lax{\mathrel{\copy\laxbox}}

\def\pa    {\ifmmode {\psi} \else {$\psi$}\fi}
\def\rPpm  {\ifmmode {r_{\Ro,\To}} \else {$r_{Ro,\To}$}\fi}

\def\vlsr  {\ifmmode {v}\else {$v$}\fi}
\def\vhelio{\ifmmode {v_{Helio}}\else {$v_{Helio}$}\fi}
\def\delV  {\ifmmode {\Delta v}\else {$\Delta v$}\fi}
\def\sigV  {\ifmmode {\sigma_v}\else {$\sigma_v$}\fi}

\def\ura   {\ifmmode {\mu_\alpha}\else {$\mu_\alpha$}\fi}
\def\udec  {\ifmmode {\mu_\delta}\else {$\mu_\delta$}\fi}

\def\ul    {\ifmmode {\mu_l}\else {$\mu_l$}\fi}
\def\ub    {\ifmmode {\mu_b}\else {$\mu_b$}\fi}

\def\uml   {\ifmmode {v_{gr}}\else {$v_{gr}$}\fi}
\def\umb   {\ifmmode {v_b}\else {$v_b$}\fi}
\def\vsrad {\ifmmode {v_{rad}}\else {$v_{rad}$}\fi}

\def\upl   {\ifmmode {v^p_{gr}}\else {$v^p_{gr}$}\fi}
\def\upb   {\ifmmode {v^p_b}\else {$v^p_b$}\fi}
\def\vprad {\ifmmode {v^p_{rad}}\else {$v^p_{rad}$}\fi}

\def\Vo    {\ifmmode {V^{Std}_\odot}\else {$V^{Std}_\odot$}\fi}
\def\Uo    {\ifmmode {U^{Std}_\odot}\else {$U^{Std}_\odot$}\fi}
\def\Wo    {\ifmmode {W^{Std}_\odot}\else {$W^{Std}_\odot$}\fi}
\def\VH    {\ifmmode {V^H_\odot}\else {$V^H_\odot$}\fi}
\def\UH    {\ifmmode {U^H_\odot}\else {$U^H_\odot$}\fi}
\def\WH    {\ifmmode {W^H_\odot}\else {$W^H_\odot$}\fi}
\def\V     {\ifmmode {V_\odot}\else {$V_\odot$}\fi}
\def\U     {\ifmmode {U_\odot}\else {$U_\odot$}\fi}
\def\W     {\ifmmode {W_\odot}\else {$W_\odot$}\fi}

\def\Vs    {\ifmmode {V_s}\else {$V_s$}\fi}
\def\Us    {\ifmmode {U_s}\else {$U_s$}\fi}
\def\Ws    {\ifmmode {W_s}\else {$W_s$}\fi}

\def\Vsbar {\ifmmode {\overline{V_s}}\else {$\overline{V_s}$}\fi}
\def\Usbar {\ifmmode {\overline{U_s}}\else {$\overline{U_s}$}\fi}
\def\Wsbar {\ifmmode {\overline{W_s}}\else {$\overline{W_s}$}\fi}

\def\aone  {\ifmmode {a_1}\else {$a_1$}\fi}
\def\atwo  {\ifmmode {a_2}\else {$a_2$}\fi}
\def\athr  {\ifmmode {a_3}\else {$a_3$}\fi}

\def\pars  {\ifmmode{\pi_s}\else{$\pi_s$}\fi}

\def\Ts    {\ifmmode{\Theta_s}\else{$\Theta_s$}\fi}

\def\To    {\ifmmode{\Theta_0}\else{$\Theta_0$}\fi}
\def\T     {\ifmmode{\Theta}\else{$\Theta$}\fi}
\def\Ro    {\ifmmode{R_0}\else{$R_0$}\fi}

\def\Tdot  {\ifmmode{\dot{\Theta}}\else{$\dot{\Theta}$}\fi}
\def\Tddot {\ifmmode{\ddot{\Theta}}\else{$\ddot{\Theta}$}\fi}

\def\lbv     {\ifmmode {(l,b,v)}\else{$(l,b,v)$}\fi}
\def\lv      {\ifmmode {(l,v)}\else{$(l,v)$}\fi}
\def\lvS     {\ifmmode {(l,v)_{\rm src}}\else{$(l,v)_{\rm src}$}\fi}
\def\lbvS    {\ifmmode {(l,b,v)_{\rm src}}\else{$(l,b,v)_{\rm src}$}\fi}
\def\lbvA    {\ifmmode {(l,b,v)_{\rm arm}}\else{$(l,b,v)_{\rm arm}$}\fi}
\def\lbvRBD  {\ifmmode {(l,b,v,R,\beta,d)}\else{$(l,b,v,R,\beta,d)$}\fi}

\def\Nbins   {\ifmmode{N_{\rm bins}}\else{$N_{\rm bins}$}\fi}
\def\DelD    {\ifmmode{\Delta d}\else{$\Delta d$}\fi}

\def\Rtp     {\ifmmode{R_{tp}}\else{$R_{tp}$}\fi}
\def\Vtp     {\ifmmode{V_{tp}}\else{$V_{tp}$}\fi}
\def\Vx      {\ifmmode{V_x}\else{$V_x$}\fi}

\slugcomment{2016 July 20}
\shorttitle{The Milky Way's Rotation Speed} 
\shortauthors{Reid \& Dame}

\begin{document}

\title{ON THE ROTATION SPEED OF THE MILKY WAY DETERMINED FROM \HI\ EMISSION 
      }   

\author{M. J. Reid\altaffilmark{1} \& T. M. Dame\altaffilmark{1}
       }

\altaffiltext{1}{Harvard-Smithsonian Center for
   Astrophysics, 60 Garden Street, Cambridge, MA 02138, USA}

\begin{abstract}
The circular rotation speed of the Milky Way at the solar radius, \To,
has been estimated to be 220 \kms\ by fitting the maximum velocity of 
\HI\ emission as a function of Galactic longitude.  This result is in 
tension with a recent estimate of $\To=240$ \kms, based on VLBI parallaxes 
and proper motions from the BeSSeL and VERA surveys for large numbers of 
high-mass star forming regions across the Milky Way.  We find that the 
rotation curve best fitted to the VLBI data is slightly curved, and that 
this curvature results in a biased estimate of \To\ from the \HI\ data 
when a flat rotation curve is assumed.  This relieves the tension between 
the methods and favors $\To=240$ \kms.
\end{abstract}


\section{Introduction} \label{sect:introduction}

The rotation speed of the Milky Way, \To, at the distance of the Sun 
from its center, \Ro, is a fundamental parameter for both Galactic and 
Local Group dynamics.  
The value of \To\ is needed to convert velocities from the 
observed Heliocentric frame to a Galactocentric frame, and it is key to 
estimating the mass of the Milky Way.  One can find estimates of \To\ 
ranging from about 170 to 270 \kms\ in the literature over the past
several decades.  In 1985, the IAU recommended $\To=220$ \kms\  
\citep{Kerr:86}, in part based on a particularly compelling analysis 
by \citet{Gunn:79} (hereafter, GKT) of the maximum observed velocity 
of \HI\ emission as a function of Galactic longitude measured in 
the inner Milky Way.
\footnote[2]{
The same authors \citep{Knapp:78} also estimated \To\ by modeling the most 
extreme H~{\tiny I} velocities in the outer Galaxy, but subsequent work 
\citep[\eg][]{Jackson:85} demonstrated that the extreme velocities differed between 
the first and fourth quadrants, as did the radial scale length of H~{\tiny I} 
in the outer Galaxy, a key input parameter in this analysis.  \citet{Knapp:88} 
ultimately concluded that the outer Galaxy data alone only constrain \To\ 
to a fairly broad range from 210 to 250 \kms.}

Recent measurements of large numbers of trigonometric parallaxes for 
masers associated with high-mass star forming regions with Very Long
Baseline Interferometry (VLBI) have provided
full 3-dimensional velocities along with ``gold standard'' distances
to sources across large portions of the Milky Way \citep{Honma:12,Reid:14}.   
Modeling the most recent data suggests a value of \To\ of $240\pm8$ \kms\ 
\citep{Reid:14}, significantly greater than the IAU recommendation.  
Independently, the {\it apparent} motion of \SgrA, assumed to be the reflex 
of the Solar orbit about the Galactic center, yields a similar value for
\To, provided $\Ro\approx8.3$ kpc \citep{Ghez:08,Gillessen:09,Do:13,Reid:14}.

Of course, other estimates of \To\ can be found in the recent literature
In particular, \citet{Bovy:12} argue for $\To\approx220$ \kms,
coupled with a very large value of the Sun's peculiar motion in the
direction of Galactic rotation of $\V=26$ \kms. ($\V=15$ \kms\ 
is used to define Local Standard of Rest (LSR) velocities.)  However, 
adopting a larger value for \V\ would not relieve the tension between the 
\HI\  estimate of $\To=220$ \kms\ and the VLBI parallax and proper motion 
estimate of $\To=240$ \kms, since it would lower both estimates.  

In this paper, we explore the method of estimating \To\ from \HI\ emission 
spectra in order to resolve this tension.   

\section{\HI\ Tangent-point Emission} \label{sect:HI}
 
\begin{figure}[ht]
\epsscale{0.65} 
\plotone{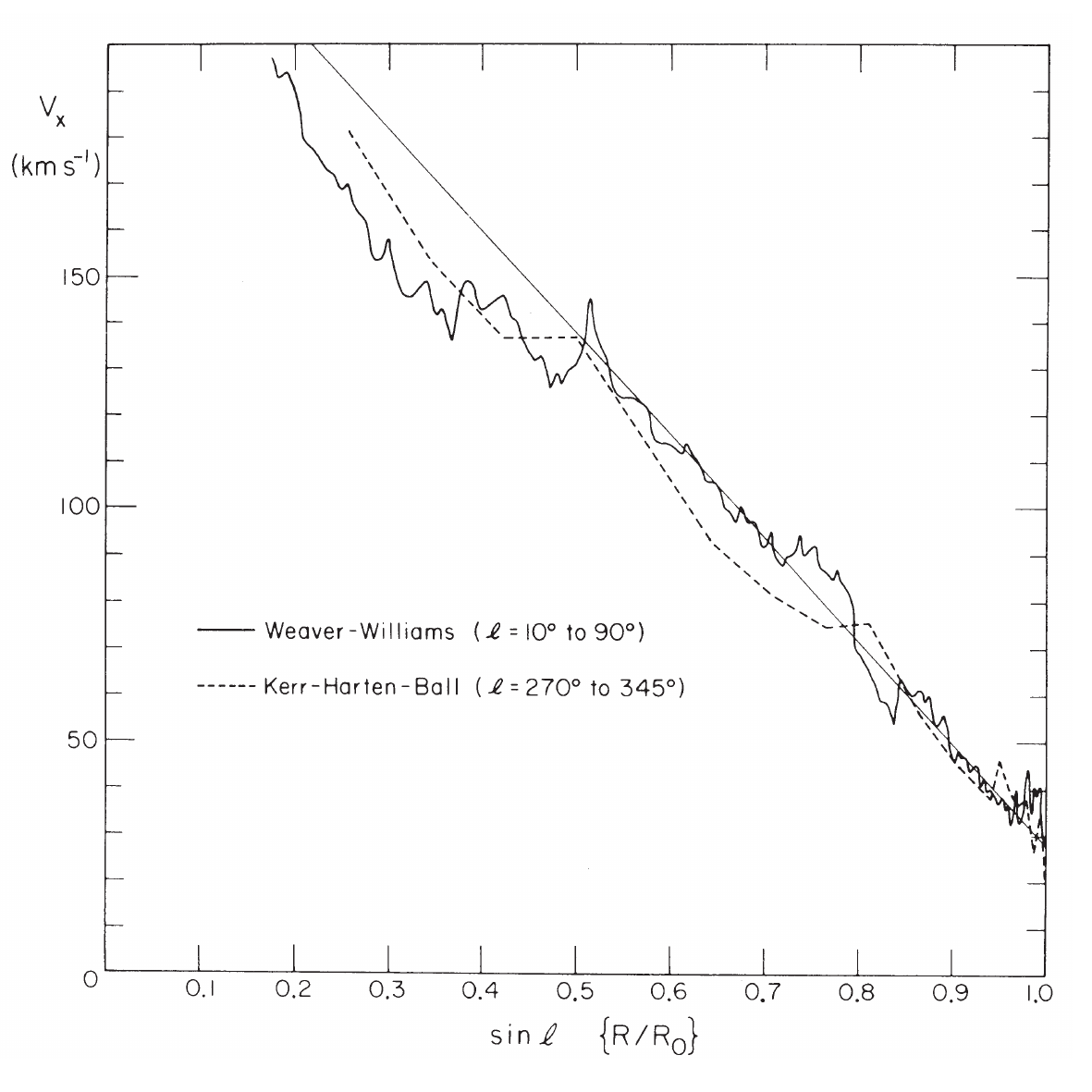}
\caption{\footnotesize
The maximum magnitude of LSR velocity, \Vx, for H~{\tiny I} emission versus sine of Galactic
logitude, reproduced from \citet{Gunn:79}.  Their linear fit over the
range of $0.5 < \sin{l} < 1$ to the first quadrant data gives a slope 
and a corresponding estimate of \To\ of 220 \kms.
        }
\label{fig:GKT_fig1}
\end{figure}

In Fig.~\ref{fig:GKT_fig1}, we reproduce the plot of the maximum LSR 
velocity of \HI\ emission as a function of the sine of the Galactic 
longitude, $l$, from GKT.  For any given longitude, the maximum positive 
velocity in the first quadrant (and maximum negative velocity in the
fourth quadrant) should occur at the tangent point. 
Along a ray from the Sun at any longitude, 
the tangent point is the closest point to the Galactic center, and at this 
point the observed LSR velocity is given by
$$\Vtp = \T(\Rtp) - \To \sin{l}~~,\eqno(1)$$
where $\T(\Rtp)$ is the circular rotation speed at the tangent point
radius from the Galactic center, $\Rtp=\Ro \sin{l}$.  GKT show traces of
the magnitude of the maximum \HI\ velocity, \Vx, in the first and fourth 
Galactic quadrants.   They fitted a straight line to the first quadrant trace
over the range $0.5 < \sin{l} < 1.0$, yielding $\Vx=28 + 220(1-\sin{l})$.
Note that at the tangent point for $\sin{l}<0.5$, radii are $\lax4.2$ kpc
and one expects significant non-circular motions for gas in the vicinity of 
the Galactic bar.

The slope of the fitted line gives a direct estimate of \To, provided the
rotation curve of the Milky Way is flat between $\Ro/2$ and $\Ro$.
GKT give a formal error on the slope of 3 \kms, but add that a realistic
uncertainty is 10 \kms, reflecting the possibility of a slight slope
in the rotation curve.  In particular, for a power-law form of the 
rotation curve, they rule out the possibility of $\To=240$ \kms, 
as it would require a rising rotation curve with an unrealistically 
large $\T(3\Ro)=280$ \kms.

Over the years there has been discussion in the literature of the 
methods used to estimate \Vx\ from \HI\ spectra.   GKT define \Vx\ 
as the extreme velocity of emission brighter than 1.0 K, and treat
the non-zero intercept (28 \kms) as owing to turbulent broadening.  
Of course, this value is sensitive to the method of estimating \Vx.
\citet{McClure-Griffiths:07} point out that \HI\ absorption against
continuum emission from \HII\ regions can affect estimates of \Vx,
favoring interferometric spectra with high angular resolution, which
allow editing of absorbed spectra.  Also, instead of using a simple
brightness threshold, they fit multiple components to the spectra to
estimate \Vx.   However, while these improvements suggest a slightly  
rising rotation curve, their estimates of \To\ still fall well below
240 \kms, whether using the \Vx\ method (see their Fig. 7) or re-scaling
the forms of the well-cited rotation curves of \citet{Burton:78} or 
\citet{Brand:93} (see their Fig. 8).

\section{The ``Universal'' Rotation Curve} \label{sect:universal}
 
\begin{figure}[ht]
\epsscale{0.65} 
\plotone{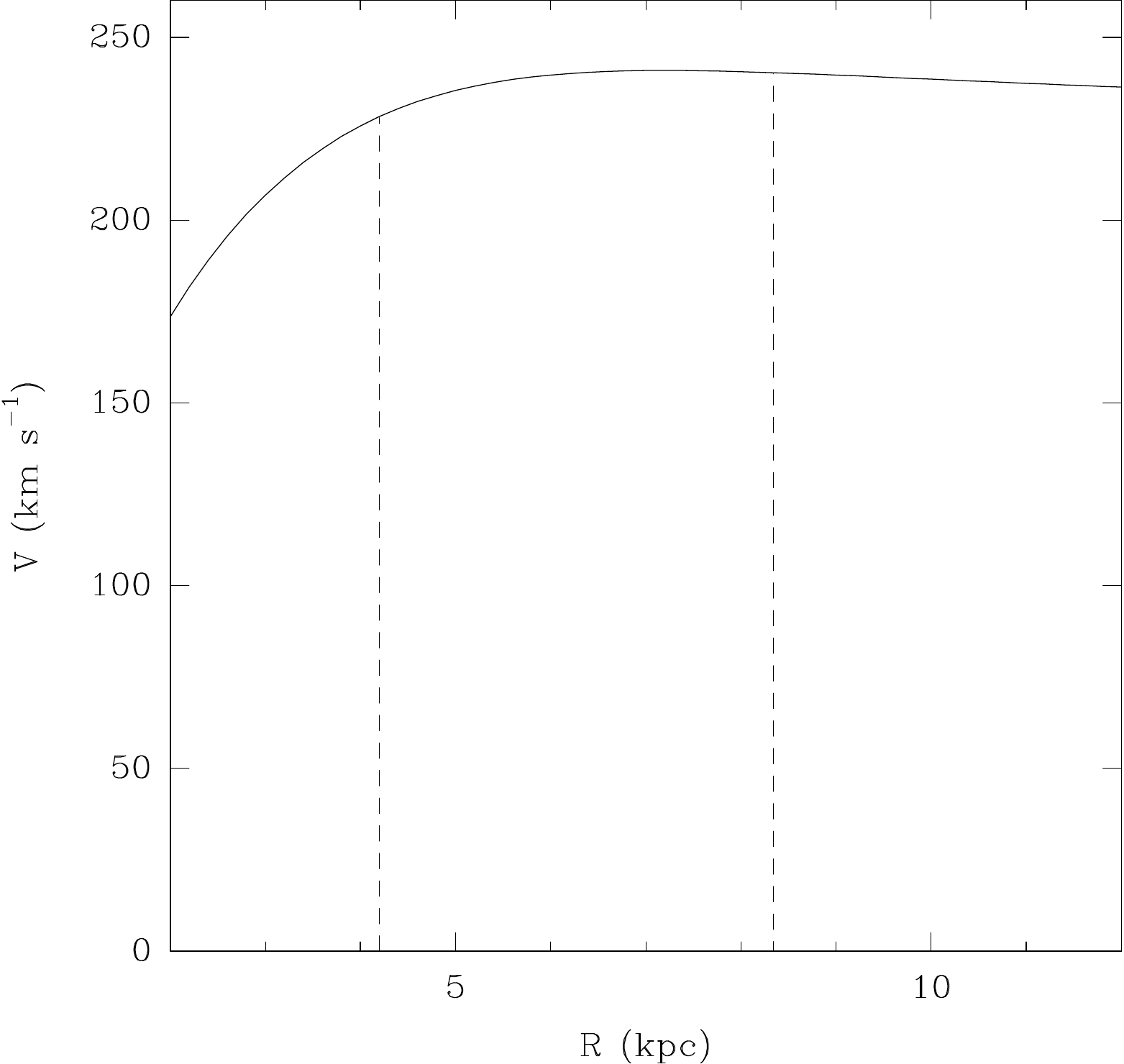}
\caption{\footnotesize
A rotation curve fitted to 80 VLBI parallaxes and proper motions of
high-mass star forming regions by \citet{Reid:14}, using the ``Universal'' 
rotation curve formulation of \citet{Persic:96}.
It rises to 228 \kms\ at 4.2 kpc ($\approx \Ro/2$), peaks at 241 \kms\ 
at $R=7.2$ kpc, and then declines very slowly at greater values of $R$.
Dashed lines enclose the range of $\Ro/2$ to \Ro\ used by GKT in fitting
\Vx\ versus $1-\sin{l}$ to estimate \To\ from H~{\tiny I} emission in the
inner Galaxy.
        }
\label{fig:Univ}
\end{figure}

VLBI parallaxes and proper motions, from the Bar and Spiral Structure 
Legacy (BeSSeL) Survey and the VLBI Exploration of 
Radio Astrometry (VERA) observations of molecular masers associated 
with high-mass star forming regions, should be excellent 
tracers of Galactic dynamics.  Eighty sources from the combined  
surveys were modeled with varous forms for the rotation curve
by \citet{Reid:14}.  The best fitting model, shown in Fig.~\ref{fig:Univ}, 
used a ``Universal'' rotation curve of \citet{Persic:96}.  This formulation 
is well motivated by observations of large numbers of spiral galaxies.  
(Specifically, the curve is for an $\L_*$ galaxy ($\beta=0.72$), with 
$V(R_{opt})=241$ \kms, $R_{opt}=0.90\Ro$ and $\Ro=8.34$ kpc.)   
Note that the fitted Universal rotation curve is nearly flat between 
about 5.5 and 8.0 kpc, but curves downward inside of 5.5 kpc.  At 4.2 kpc 
($\approx\Ro/2$) the rotation speed is 228 \kms, it peaks at 241 \kms\ 
at 7.2 kpc, and it declines very slightly to 240 \kms\ just past \Ro\ (assumed
to be 8.34 kpc).

\begin{figure}[ht]
\epsscale{1.0} 
\plotone{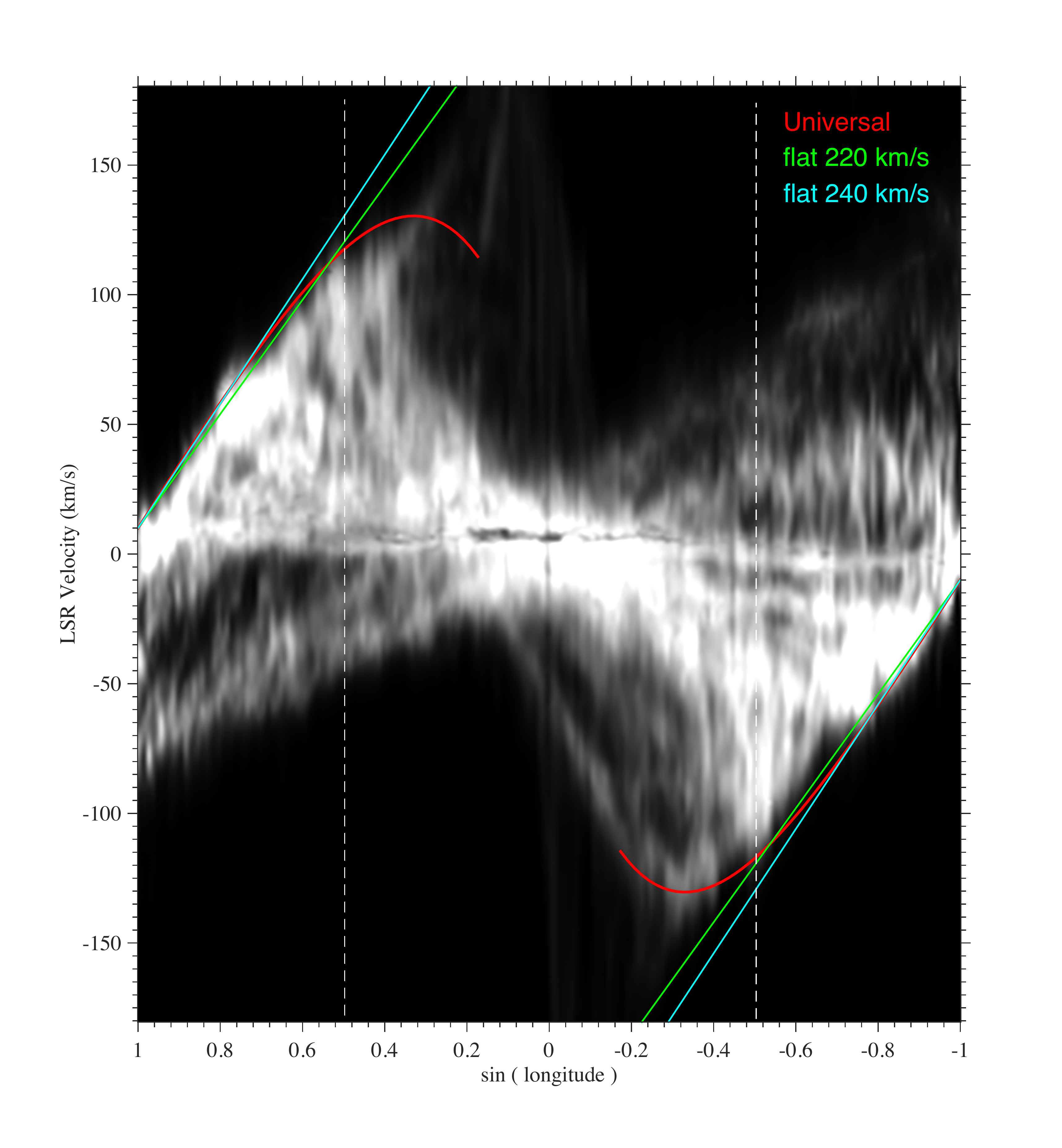}
\caption{\footnotesize
The maximum LSR velocity for three rotation curves versus the sine of 
Galactic logitude, superposed on H~{\tiny I} emission ({\it grayscale}) from 
the Leiden/Argentine/Bonn Galactic H~{\tiny I} Survey \citep{Kalberla:05}.  
Plotted are maximum velocites for flat rotation curves with $\To=220$ 
and $240$ \kms\ in {\it green} and {\it cyan} lines, respectively.  
Also plotted in {\it red} is the maximum velocity for the ``Universal'' 
rotation curve from \citet{Reid:14}.  All three curves assume a turbulent
broadening of 10 \kms, appropriate for the grayscale stretch.
{\it Dashed lines} indicate the range of $\sin{l}=0.5$ at \Ro/2 to 
$\sin{l}=1$ at \Ro\ used by GKT when fitting \Vx\ versus $1-\sin{l}$ to 
estimate \To\ from H~{\tiny I} emission in the inner Galaxy.  
        }
\label{fig:HI}
\end{figure}

As Fig.~\ref{fig:HI} shows, \Vx\ predicted by the Universal rotation curve 
matches the highest velocity of \HI\ emission very well over 
most of the inner Milky Way, and at $\sin{l} < 0.5$ it provides a
much better fit than for flat rotation curves with \To\ of 220 or 240 
\kms.  However, note that over the longitude range fitted by GKT
($0.5 < \sin{l} < 1.0$), the Universal curve is roughly linear in \Vx\ 
and can be approximated by a flat curve of 220 \kms.   
{\it This holds even though the Universal rotation 
curve has a circular velocity of 240 \kms at the solar radius and
an average velocity of 238 \kms\ over the fitted range.}

\begin{figure}[ht]
\epsscale{0.65} 
\plotone{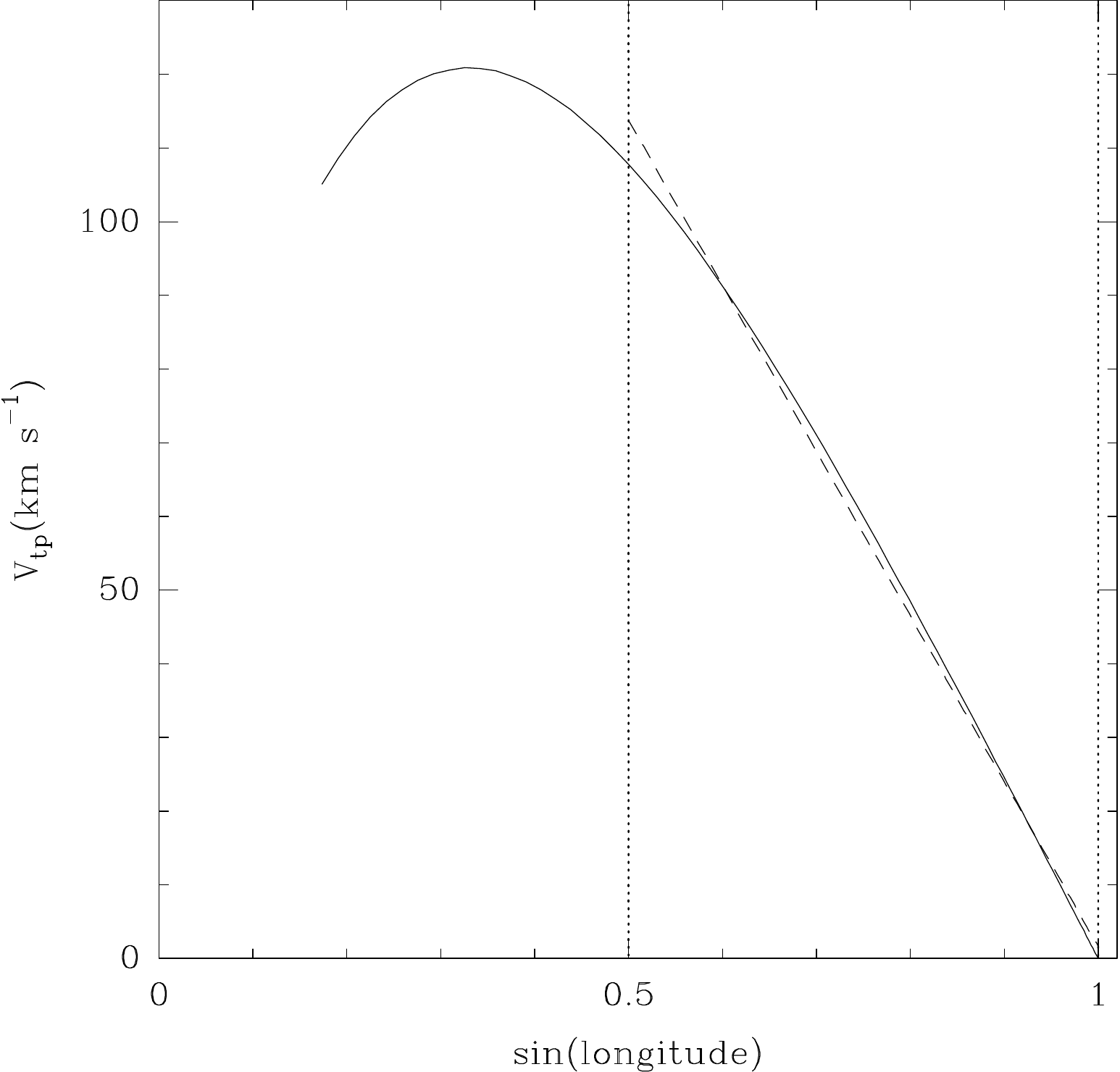}
\caption{\footnotesize
The tangent-point (maximum) velocity for the ``Universal'' rotation curve 
of \citet{Reid:14} versus the sine of Galactic longitude ({\it solid line}).
A linear fit over the range indicated by the {\it dotted lines} of 
$0.5 < \sin{l} < 1$ is shown with the {\it dashed line}.
The slope of the linear fit is 220 \kms, even though the rotation curve
never falls below 228 \kms\ and has an average value of 238 \kms\ over the 
fitted range. 
        }
\label{fig:tp_Univ}
\end{figure}

To further demonstrate this point, in Fig.~\ref{fig:tp_Univ} we fit 
\Vtp\ predicted by the Universal curve with a straight line over the 
GKT range of $\sin{l}$ and find a slope of 224 \kms, consistent with 
the GKT result.  Alternatively, simply defining a straight line by the 
rotation curve values at the endpoints of their fitting range yields a 
similar value of 220 \kms.

We conclude that a rotation curve that rises from 228 \kms\ at 4.2 kpc to
241 \kms\ at 7.2 kpc and then flattens, as does the Universal rotation curve 
fitted to the VLBI data, gives a biased estimate of \To\ that is low by about 
20 \kms, when fitting a straight line to tangent-point velocities
between radii of $\Ro/2$ and \Ro.  This resolves the apparent tension 
between the two methods of estimating \To\ and favors a value of 240 \kms.

\vskip 0.5truein\noindent 
{\it Facilities:}  \facility{VLBA}, \facility{VERA}, \facility{EVN}


\begin{thebibliography}{}
\bibitem[Bovy \etal(2012)]{Bovy:12}
         Bovy, J., Prieto, C. A., Beers, T. C. \etal 2012, \apj, 759, 131
\bibitem[Brand \& Blitz(1993)]{Brand:93}
         Brand, J. \& Blitz, L. 1993, \aaps, 275, 67
\bibitem[Burton \& Gordon(1978)]{Burton:78}
         Burton, W. B. \& Gordon, M. A. 1978, \aaps, 63, 7
\bibitem[Do \etal(2013)]{Do:13}
         Do, T. Martinez, G. D., Yelda, S. \etal 2013, \apjl, 779, L6
\bibitem[Ghez \etal(2008)]{Ghez:08}
         Ghez, A. M., Salim, S. Weinberg, N. N. \etal 2008, \apj, 689, 1044
\bibitem[Gillessen \etal(2009)]{Gillessen:09}
         Gillessen, S., Eisenhauer, F., Trippe, S. \etal 2009, 
         \apj, 692, 1075
\bibitem[Gunn, Knapp \& Tremaine(1979)]{Gunn:79}
         Gunn, J. E., Knapp, G. R. \& Tremaine, S. D. 1979, 
         \aj, 84, 1181
\bibitem[Honma \etal(2012)]{Honma:12}
         Honma, M., Nagayama, T., Ando, K, \etal 2012, \pasj, 64, 136
\bibitem[Jackson(1985)]{Jackson:85}
         Jackson, P. D. 1985, in The Milky Way, IAUS 106, 
         eds. H. van Woerden, R. J. Allen \& W. B. Burton, p.179
\bibitem[Kalberla \etal(2005)]{Kalberla:05}
         Kalberla, P. M. W., Burton, W. B., Hartmann, Dap, \etal 2005, 
         \aaps, 440, 775
\bibitem[Kerr \& Lynden-Bell(1986)]{Kerr:86}
         Kerr, F. J. \& Lynden-Bell, D. 1986, \mnras, 221, 1023
\bibitem[Knapp(1988)]{Knapp:88}
         Knapp, G. R. 1988, in Lecture Notes in Physics, eds. L. Blitz \& 
         F. J. Lockman (Springer-Verlag, Berlin), 306, 3
\bibitem[Knapp, Tremaine \& Gunn(1978)]{Knapp:78}
         Knapp, G. R., Tremaine, S. D. \& Gunn, J. E. 1978, \aj, 83, 1585
\bibitem[McClure-Griffiths \& Dickey(2007)]{McClure-Griffiths:07}
         McClure-Griffiths, N. M. \& Dickey, J. M. 2007, \apj, 671, 427
\bibitem[Persic, Salucci \& Stell(1996)]{Persic:96}
         Persic, M., Salucci, P. \& Stel, F. 1996, \mnras, 281, 27
\bibitem[Reid \etal(2014)]{Reid:14}
         Reid, M. J., Menten, K. M., Brunthaler, A. \etal\ 2014, \apj, 783, 130
\end{thebibliography}
\end{document}